\documentclass[12pt]{article}

\newcommand{\be}{\begin{equation}}
\newcommand{\ee}{\end{equation}}
\newcommand{\ben}{\begin{eqnarray}}
\newcommand{\een}{\end{eqnarray}}
\newcommand{\n}{\label}
\newcommand{\no}{\noindent}
\newcommand{\la}{\lambda}

\usepackage{graphicx}

\begin{document}

\title{Unified phantom  cosmologies}
\author{ \large Luis P. Chimento$^1$\footnote{chimento@df.uba.ar} \addtocounter{footnote}{5}  and Ruth Lazkoz $^2$\footnote{wtplasar@lg.ehu.es}\\
{\it \small Dpto. de F\'\i sica, Facultad de Ciencias Exactas y Naturales, }
\\{\it \small Universidad de Buenos Aires, Ciudad Universitaria,}\\
{\it \small  Pabell\'on I, 1428 Buenos Aires, Argentina}\\
{\it  $^1$ \small Fisika Teorikoa,  Zientzia eta Teknologia Fakultatea, }\\
{\it \small Euskal Herriko Unibertsitatea,}\\
{\it \small  644 Posta Kutxatila, 48080 Bilbao, Spain}}
\maketitle

\begin{abstract}

We present a general algorithm based on the concept of form-invariance which
can be used for  generating phantom cosmologies. It involves  linear
transformations  between the kinetic energy and the potential of the
scalar field, and transforms solutions of the Einstein-Klein-Gordon equations
which preserve the weak energy condition into others which violate it, while
keeping the energy density of the field positive. All
known solutions representing phantom cosmologies are unified by this procedure. 
Using the general algorithm we obtain those solutions and show the relations between them. In addition,
the scale factors of the product and seed solutions are related by a
generalization of the well-known $a\to a^{-1}$ duality.

\end{abstract}

\newpage

\section{Introduction}
Recently obtained astrophysical information ranging from high redshift surveys of supernovae
to Wilkinson Microwave Anisotropy Probe (WMAP) observations, indicates that a considerable amount of the total energy
of our universe might correspond to some dark energy   which would be currently producing
inflation thanks to a large and negative
pressure. 

Such exotic fluids may be framed in theories with matter fields that violate the
weak energy condition
\cite{caldwell}. These models were dubbed phantom
cosmologies, and their study represents 
a currently active area of research in theoretical cosmology. Although phantom
cosmologies have
been investigated from different perspectives, here we will only be concerned with particular issues
related with analytical properties \cite{earlier}.

At present,
the theoretical understanding of the subject is limited, but at least one can rely on the motivation
for phantom matter which is provided by string
theory \cite{str}. Precisely, the idea that the origin
of dark energy should be searched within string theory or another fundamental
theory has been recently reinforced by
the discovery that the holographic principle cannot be
used to tell whether dark energy is present or not \cite{holography}.

One interesting aspect of phantom matter is that it 
might make the universe end up in a kind of
singularity \cite{bigrip} characterized by  divergences in the scale factor
$a$, the Hubble factor $H$ and its  time-derivative $\dot H$. This possibility has not only been shown to exist in general relativity \cite{bigripgen}, but also in some of its
generalizations like braneworld cosmology \cite{bigripbrane} or scalar-tensor theories of gravity \cite{bigripst}.

Another attractive idea put forward recently in connection with phantom fields is that they can be used constructing bouncing universes \cite{bouncing}.

In this paper we present an algorithm for generating  phantom cosmologies with
Friedmann-Robertson-Walker (FRW) geometries. Interestingly, the method unifies
others given in \cite{letter} and \cite{Lidsey}. In fact, here we will prove
that the triality  transformation implemented in \cite{Lidsey} which chains  inflationary, cyclic  
and phantom cosmologies is a form-invariance transformation; moreover, such transformation is itself a composition of three simpler form-invariance symmetry operations. Thus, that procedure is an application of general form-invariance
transformations, which have been  used successfully in \cite{letter} and
\cite{nuestros} for generating new exact solutions.

Once we discuss the main general features of those transformations we
concentrate on the linear ones. After that, we show there exist four
transformations  which preserve the positivity of the energy density, 
and we demonstrate they can be used for generating phantom cosmologies starting
from standard scalar field universes. We will be mainly be concerned with universes represented by scalar field solutions to the Einstein equations, and as such  the energy density will be 
sum of a kinetic and a potential term. Now, in the usual definition of the  energy density  of a scalar field the kinetic term
is taken to be positive, but we will relax this hypothesis and will also consider the possibility that it is negative.
The transformations we consider map between them solutions  constructed by
taking the two possible signs of the kinetic term. This will result in
two possible ways of transforming those scalar field cosmologies
into phantom ones. We  will discussed this in detail and we will put some accent
on comparison with related results which appeared earlier in the literature.   
Finally, we will draw our main conclusions.

\section{The form invariance symmetry}

Our starting point are perfect fluid  FRW spacetimes with flat sections given by the line element
\begin{equation}
\label{rw}
ds^{2}=-dt^{2}+a^{2}(t)\left(dx^{2}+dy^{2}+dz^{2}\right).
\end{equation}
Note that, even in such a simple setting as this, many questions remain open yet.
The Einstein equations are

\begin{equation}
\label{00}
3H^{2}=\rho,
\end{equation} 

\be
\n{co}
\dot\rho+3H(\rho+p)=0,
\ee

\no where $\rho$ and $p$ are respectively the energy density and the pressure of the fluid, and $H\equiv \dot a/a$. For
a different perfect fluid with energy density $\bar\rho$ and pressure $\bar p$
the Einstein equations will take the form

\begin{equation}
\label{00b}
3\bar H^{2}=\bar\rho,
\end{equation} 

\be
\n{cob}
\dot{\bar\rho}+3\bar H(\bar\rho+\bar p)=0.
\ee
The system of equations
(\ref{00b})-(\ref{cob}) is known to admit form-invariance transformations \cite{luis}, which means there exist symmetry transformations which  relate it to the system (\ref{00})-(\ref{co}). Those transformations are given by

\be
\n{tr}
\bar\rho=\bar\rho(\rho),
\ee

\be
\n{th}
\bar H=\left(\frac{\bar\rho}{\rho}\right)^{1/2}H,
\ee

\be
\n{tp}
\bar
p=-\bar\rho+
\left(\frac{\rho}{\bar\rho}\right)^{1/2}(\rho+p)\frac{d\bar\rho}{d\rho},
\ee

\no where $\bar\rho=\bar\rho(\rho)$ is an invertible function.  The symmetry
transformations (\ref{tr})-(\ref{tp}) define a Lie group for any function
$\bar\rho(\rho)$, which can connect scenarios with different evolutions. This will be discussed in the following sections.

Nevertheless,  it is worth making a pair of clarifications before we move on. First,  form-invariance transformations
are defined without imposing any restriction on the fluid. Second, they relate solutions
to different equations, unlike Lie point symmetry transformations, which relate
different solutions to the same equation. Therefore, form-invariance may be viewed as 
an uncommon yet useful equivalence concept.

The implications of the symmetry transformations will be better understood if one is able to say how  they change
relevant physical parameters. We will look into this matter from a a general perspective and also  from that of well-known families of solutions. Under the symmetry transformations (\ref{tr})-(\ref{tp}), the deceleration factor $q$, which by definition is
\begin{equation}q=-H^{-2}\frac{\ddot{a}}{a},\end{equation} changes according to 

\be
\n{tq}
\bar q+1=\left(\frac{\rho}{\bar\rho}\right)^{3/2}\frac{d\bar\rho}{d\rho}\,
(q+1).
\ee

If we consider
perfect fluids with equations of state

\be
\n{ece}
p=\left(\gamma-1\right)\rho, \qquad
\bar p=\left(\bar\gamma-1\right)\bar\rho,
\ee

\no respectively, then the barotropic index $\gamma$ will become $\bar\gamma$
and  the link between them will be

\be
\n{tg}
\bar\gamma=\left(\frac{\rho}{\bar\rho}\right)^{3/2}\frac{d\bar\rho}{d\rho}\,
\gamma.
\ee

We have seen that the form-invariance transformations we are concerned with 
require as only input    the relation between
the energy densities of the two fluids. Thus, according to that,
we may consider the linear transformation between
$H$, $\rho$ and $p$ generated by
\be
\n{tn}
\bar\rho=n^2\rho, 
\ee

\no where $n$ is a real number. Even though the latter a simple choice, it has got much
in store. The transformed barotropic index and  deceleration factor will read

\be
\n{gs}
\bar\gamma=\frac{\gamma}{n}.
\ee
and
\be
\n{tqs}
\bar q=-1+\frac{q+1}{n}.
\ee
\no There will be  inflation ($\bar q<0)$ for any $n>(q+1)$ if $n>0$
or for any $n<(q+1)$ if $n<0$.
The first case was studied in full detail in \cite{luis},
whereas the second one was partially investigated in
\cite{letter}. In this case we are going to devote ourselves to  the second case.

Finally, for the linear relation (\ref{tn}), the transformation rule
(\ref{tr})-(\ref{tp}) between the physical quantities gives

\begin{eqnarray}
\left(\begin{array}{c}
\bar \rho\\
\bar H\\
 \bar p\\
\end{array}\right)=
\left(\begin{array}{ccc}
n^2 & 0 & 0\\
0 & n & 0\\
 n-n^2& 0 & n
\end{array}\right)\left(\begin{array}{c}
\rho\\
H\\
p\\
\end{array}\right).
\n{m}
\end{eqnarray}
This means we are actually dealing with a linear transformation acting on
the variables $(\rho,H,p)$, which arises because
of the particular choice  (\ref{tn}). In general other form-invariance transformations will not
be linear.
Now, the second row in Eq. (\ref{m}) tell us that

\be
\n{thg}
\bar H=nH,
\ee

\no which leads to

\be
\n{a-}
\bar a=a^n.
\ee

The insight gained in this section will be used in the remainder with the purpose of  generating phantom cosmologies. 

\section{Phantom cosmologies and duality}

When solving the Einstein equations of flat FRW models (\ref{00}) and
(\ref{co}), one looks for solutions of a system of equations in which $a$ does
not appear explicitly; $H$, $\rho$ and $p$ are the true variables of the
problem, and mathematically speaking the most natural choice is making
assumptions on them, instead of the scale factor. 

Here we will consider models such that the Hubble factor is linear in $1/t$.
They correspond to power-law evolutions and have a  constant barotropic index.
Thus, because of the equations of state (\ref{ece}) chosen earlier, we will
have
\begin{eqnarray}
\n{hl}
\bar H=\frac{2}{3\bar\gamma t}\\  H=\frac{2}{3\gamma t}\label{seed},
\end{eqnarray}
where $\bar\gamma$ and $\gamma$ are the bariotropic index of the two fluids.
The scale factor of the seed solution
associated with (\ref{seed}) reads
\begin{equation}
a(t)=(\pm \,t)^{2/3\gamma} \label{powerlaw}.
\end{equation}

\no From Eq. (\ref{seed}), we have expanding solutions for $\gamma t>0$ and
contracting ones for $\gamma t<0$. In order to be able to carry out the
following discussion for a positive $\gamma$, we will take the plus sign for
the $t>0$ branch of the solution and the minus sign for the $t<0$ one. The
conection between these two time regions is made throughout a time reversal
symmetry. Since this transformation does not change the Einstein equation
(\ref{00})-(\ref{co}), it is a form-invariance transformation. Under that sign
choice the two branches will be mirror images of each other. Now, replacing
Eqs. (\ref{hl})-(\ref{seed}) in (\ref{thg}) and integrating one arrives at

\be
\n{pw}
|t|^{2/3\bar\gamma}=|t|^{2n/3\gamma}.
\ee

Since in the $n<0$ case one has $a\to a^{-\vert n\vert}$ the transformation
may be viewed as an extension of the well-known $a\to a^{-1}$ duality given in
our paper \cite{letter}, and which was profusely studied in the pre-big bang
scenario \cite{prebigbang}. Now a solution of the form of (\ref{powerlaw})
with $\gamma>0$ describes for $t>0$ a standard universe that expands ever
after a big-bang at $t=0$, but the $t<0$ branch describes a contracting
solution for which $t=0$ represents a big crunch. Now, when we apply the
transformation we obtain a new solution which contracts in its $t>0$ branch,
but expands in the $t<0$ one. Precisely, the latter satisfies the definition
of a  phantom universe given in the literature because it is a expanding
solution that violates the weak energy condition. Moreover, it reaches a big
rip at finite time, which is a feature inherent to many phantom universes but
not to all of them. Thus, the phantom universe is dual to the universe
described by the $t<0$ branch of the seed solution.

\begin{figure}[t]
\label{figure:sources_2}
\begin{center}
\includegraphics[width=0.4\textwidth]{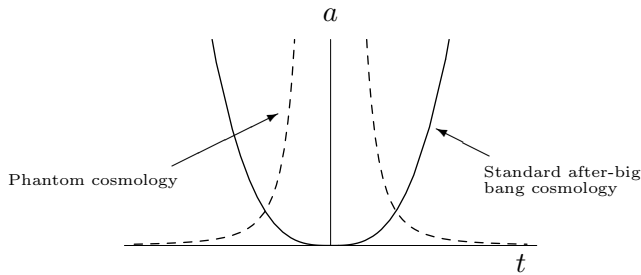}
\put(-20,26){\tiny Standard after-big}
\put(-20,20){\tiny bang cosmology}
\put(-200,23){\tiny Phantom cosmology}
\put(-81,85){\small$a$}
\put(-8,-10){\small $t$}
\put(-138,30){\vector(2,1){40}}
\put(-18,34){\vector(-2,1){20}}
\end{center}
\caption{The continuous line is a plot of the scale factor as a function of time for a standard after-big bang cosmology ($t>0$) and its pre-big bang continuation. The dashed line represents a new solution obtained by applying to the latter a generalized duality transformation  $a\to a^{-1\vert n\vert}$. The $t<0$ region of the new solution represents
a phantom cosmology reaching a big-rip at finite time.}
\end{figure}

\section{A unified generation algorithm}

In this section we turn our attention to the scalar field picture; in
particular, we study how scalar fields transform under the form invariance
transformations
considered above. To this end, we identify the corresponding energy-momentum
tensor with that of a perfect fluid and write the energy densities an
pressures  in the invariant form

\be
\n{rp}
\rho(\phi)=s\frac{\dot\phi^2}{2} + V(\phi),\quad p(\phi)=s\frac{\dot\phi^2}{2} - V(\phi),
\ee
and
\be
\n{rpt}
\bar\rho(\bar\phi)=\bar s \frac{\dot{\bar\phi}^2}{2}+ \bar V(\bar\phi),\quad\bar p(\bar\phi)=\bar s\frac{\dot{\bar\phi}^2}{2} - \bar V(\bar\phi).
\ee

\no where $s$ and $\bar s$ are two parameters of the scalar field models which
can be chosen freely. 

Since the energy density and pressure are written as a sum of two terms
(kinetic and potential energy), there is in principle the possibility of
combining them linearly so that the sums   $\rho$ and $p$ (or $\bar \rho$ and
$\bar p$) remain constant. This generates an additional group structure other
than the one introduced by the transformation (\ref{tn}). However, since we
are only interested  in changes of sign in the kinetic energy we will restrict
ourselves to the  discrete symmetries obtained in that case. Therefore,
redefining the scalar fields $\phi$ and $\bar\phi$, we can select the constant
$s$ and $\bar s$ as $s=\pm 1$ y $\bar s=\pm 1$. Thus, they determine the sign
of the kinetic terms in the theories described by Eqs. (\ref{rp}) and
(\ref{rpt}) respectively.

If we insert (\ref{rp}) and (\ref{rpt}) in (\ref{co})
and (\ref{cob}) respectively,  the original and form-invariance transformed
Einstein-Klein-Gordon equations will be obtained:

\ben
\n{kg}
\ddot{\phi}+3H\phi+\frac{dV(\phi)}{sd\phi}=0,\\
\n{kgb}
\ddot{\bar\phi}+3\bar H\bar\phi+\frac{d\bar V(\bar\phi)}{\bar sd\bar\phi}=0.
\een

\no 
 
Applying the form-invariance transformations (\ref{tr})-(\ref{tp}) and particularizing to
our choice (\ref{tn}), we find that the kinetic and potential terms transform
linearly as

\be
\n{tt}
{\dot{\bar\phi}}^2=\frac{s}{\bar s}n\dot\phi^2,
\ee

\no and
\be
\n{tv}
\bar V=ns(n-1)\frac{\dot\phi^2}{2}+n^2V.
\ee

Let us see now in more detail  how the transformation operates. Under the
definitions made above we have

\be
\n{rps}
-2\dot H=\rho+p=s{\dot\phi}^2.
\ee 
and, in consequence, our transformation (\ref{thg}) gives
\begin{equation}
\bar\rho+\bar p=n(\rho+p).
\end{equation}

 If the seed solution preserves the weak energy condition and considering
we are in the  $t<0$ region with $n<0$, then the  transformed
solution will violate it and will terminate in a big rip \cite{bigrip}
(alternatively see \cite{barrow} for big rip singularities in non-phantom
settings). We concentrate in the remainder on the $n=-1$ case, which
corresponds to $\rho\to\rho$, $H\to-H$ and $\rho+p\to-(\rho+p)$. Moreover, the
barotropic index of the associate fluid is 
\be
\gamma=(\rho+p)/\rho=-2\dot
H/3H^2=s\frac{\phi^2}{\rho}\ee
 and its transformation rule (\ref{tg}) reads $\gamma\to -\gamma$.

Due to the transformation law of the scalar field, the swap between cosmologies that
preserve the energy condition and others that violate it  can occur in two
different ways, depending on whether  it is the sign of $s$ or ${\dot \phi}^2$
that flips  in the transformation \footnote{If the sign of ${\dot \phi}^2$ is
preserved by the transformation the energy density we will behave the kinetic
energy behaves as a scalar, whereas if that sign changes we will say the
kinetic energy behaves as a pseudo-scalar.}. Combining this double possibility
with the also double choice for the character of the seed solution
($s=1$ or  $s=-1$)  we obtain, from Eqs. (\ref{tt})
and (\ref{tv}) four possible transformations. We list them in Table 1 for the particular case $n=-1$.

\begin{table}[h!]
\caption[crit]{ List of $n=-1$ transformations.}\label{crit}
\begin{center}
\begin{tabular}{ccccc}\\
\hline
\hline
\vspace{-0.3cm}
\\
$s$ &$\bar s$& $\bar \phi$ & $\bar V$ & $\bar \gamma$ \\
\hline\vspace{-0.3cm}\\
$\,1$ &$\; \;1$& $\mp i\phi$ & $\;\;\dot\phi^2+V$ & $\displaystyle-{\dot\phi^2}/{\rho}$ \\\\
$\,1$ &$ -1$& $\pm \phi$ & $\;\;\dot\phi^2+V$ & $\;\;\displaystyle{\dot\phi^2}/{\rho}$ \\\\
$-1$ &$ \;\;1$& $\pm \phi$ & $-\dot\phi^2+V$ & $\;\;\displaystyle{\dot\phi^2}/{\rho}$ \\\\
$-1$ &$- 1$& $\mp i\phi$ & $-\dot\phi^2+V$ & $\displaystyle-{\dot\phi^2}/{\rho}$ \\\\\hline
\hline
\end{tabular}
\end{center}
\end{table}

Now, further insight on how the transformation works can be gained from the
field equations. We will distinguish between conventional and non conventional
scalar fields by writing the Einstein equations using respectively  positive
or  negative  kinetic terms. For a conventional scalar field (CSF) the set of
Einstein equations is

\begin{equation}
\label{c1}
3H^{2}=\frac{1}{2}\dot\phi^{2}+V(\phi),
\end{equation} 

\begin{equation} 
\label{c2}
\dot{\rho}+3H\dot{\phi}^2=0,
\end{equation}

\no whereas for an non conventional scalar field (NCSF) the set of Einstein
equations is
                                                    
\begin{equation}
\label{n1}
3H^{2}=-\frac{1}{2}\dot\phi^{2}+V(\phi),
\end{equation} 

\begin{equation} 
\label{n2}
\dot{\rho}-3H\dot{\phi}^2=0.
\end{equation}

The first transformation in Table I makes $ \dot{\phi}^2\to-\dot\phi^2$, and $
V\to\dot\phi^2+V$, and  leaves unaltered the CSF equation set.  The second
transformation makes $\dot{\phi}^2\to\dot\phi^2$ and  $V\to\dot\phi^2+V$, so
that it turns the set of equations governing the NCSF into the set CSF. The
third transformation makes $\dot{\phi}^2\to\dot\phi^2$ and
$V\to-\dot\phi^2+V$, and it turns the set of equations governing the CSF into
the NCSF set. Finally, the fourth transformation makes
$\dot{\phi}^2\to-\dot\phi^2$, and $V\to-\dot\phi^2+V$, and leaves unaltered
the NCSF equation set.  

It is also interesting to look at the implications of the transformations from
the point of view of the barotropic index $\gamma$. A solution to the CSF set
with a real scalar field will have a positive barotropic index, and its $t>0$
and $t<0$ branches  will represent respectively expanding and contracting
universes with a null scale factor at $t=0$.  Taking a seed of that sort one
can construct phantom solutions  to the CSF set with an imaginary scalar field
(using the $s=\bar s=1$ transformation) or with a real scalar field (using the
$s=-\bar s=1$ transformation). Specifically, these phantom universes will
occupy the $t<0$ region. In contrast, a solution to the NCSF set having a
negative barotropic index with a real scalar field  and its $t>0$ and $t<0$
branches  will represent respectively contracting and expanding universes with
an infinite scale factor at $t=0$, so its $t<0$ region branch is a phantom
universe. Application of the duality transformation gives us  a new solution
such that its $t>0$ and $t<0$ branches  will represent respectively expanding
and contracting universes with a null scale factor at $t=0$. We stress here
again that if $n \ne -1$ the absolute value of the barotropic index will
change in the transformation according to $\bar\gamma=\gamma/n$.

For illustration we consider now the same families of exact solutions
considered earlier, that is, the models in which the Hubble factor is
proportional to $1/ t$. Choosing all the terms in each equation (\ref{00}) and
(\ref{kg}) with the same degree of homogeneity, the existence of solutions is
guaranteed. Hence, the corresponding scalar field has a logarithmic dependence
on time $t$ and the potential depends on the inverse square of $t$, so that

\ben
\n{ct}
\bar\phi=\frac{2}{\bar A}\ln|t|, \qquad \phi=\frac{2}{A}\ln|t|,
\een

\no and

\be
\n{pt}
\bar V=\frac{\bar V_0}{t^2},\qquad \n{ptb} V=\frac{V_0}{t^2},
\ee

\no where $\bar A$, $A$, $\bar V_0$ and $V_0$ are parameters related by the
transformation rules for the field and the potential (\ref{tt})-(\ref{tv}). In
addition, they are restricted by the respective Einstein-Klein-Gordon
equations. Then, inserting Eqs. (\ref{ct}) and (\ref{ptb}) into the Eqs.
(\ref{tt})-(\ref{tv}), we get

\be
\n{tA}
{\bar A}^2=\frac{\bar s}{ns}A^2,
\ee

\be
\n{tv0}
\bar V_0=n^2\left(\frac{2s}{A^2}+V_0\right)-\frac{2ns}{A^2},
\ee

\no while the composition of (\ref{ct}) with (\ref{pt}) gives the potentials
as  functions of the fields
\ben
\n{po}
\bar V=\bar V_0 {\mbox e}^{-\bar A\bar\phi}, \qquad
V=V_0 {\mbox e}^{-A\phi},
\een

Requiring that the Hubble factor $H$, the field $\phi$ and the potential $V$
given by (\ref{hl}), (\ref{ct}) and (\ref{pt}) respectively satisfy the
unbarred Einstein-Klein-Gordon equations (\ref{00}) and (\ref{kg}) we find
that

\ben
\n{cons}
\gamma=\frac{A^2}{3s},
\qquad V_0=\frac{2s}{A^2}\left(\frac{6s}{A^2}-1\right).
\een

\no Finally, replacing the latter in (\ref{tv0}) we arrive at the complete
transformation law for the parameter $\bar V_0$:

\be
\n{tv0f}
\bar V_0=\frac{2ns}{A^4}\left(6ns-A^2\right).
\ee

For the seed power law solution
$a(t)=|t|^{2/3\gamma}$ or $a(t)=|t|^{2s/A^2}$ (by using Eq. (\ref{cons})), the
symmetry operations of Table 1 give rise to four different kinds of solutions. We list them in Table 2.

\begin{table}[h!]
\caption[crit]{ Transformations of the power-law solutions in the  $n=-1$ case.}\label{crit}
\begin{center}
\begin{tabular}{cccccc}\\
\hline
\hline
\vspace{-0.3cm}
\\
$s$ &$\bar s$& $\bar \phi$ & $\bar a$ & $\bar A$ & ${\bar V}_0$\\
\hline\vspace{-0.3cm}\\
$1$ &$ 1$& $\displaystyle\mp \frac{2i}{A}\ln\vert t\vert$ & $|t|^{-2/A^2}$ & $\pm iA$ & 
$\displaystyle\frac{2}{A^4}(6+A^2)$\\\\
$1$ &$ -1$& $\displaystyle \frac{2}{A}\ln\vert t\vert$ & $|t|^{-2/A^2}$ & $\pm A$ & $\displaystyle\frac{2}{A^4}(6+A^2)$\\\\
$-1$ &$ 1$& $\displaystyle \frac{2}{A}\ln\vert t\vert$ & $|t|^{2/A^2}$ & $\pm A$ & $\displaystyle\frac{2}{A^4}(6-A^2)$\\\\
$-1$ &$- 1$& $\displaystyle\mp \frac{2i}{A}\ln\vert t\vert$ & $|t|^{2/A^2}$ & $\pm iA$ & $\displaystyle\frac{2}{A^4}(6-A^2)$\\\\\hline
\hline
\end{tabular}
\end{center}
\end{table}

The solution generation procedure we discussed in \cite{letter} corresponds to
the first transformation. More specifically, if we take the usual power-law
cosmologies with a constant barotropic index (the ones used above) and apply
the transformation $\bar s=-1=-s$ we get a family of power-law solutions that
are, in fact, the late time limit of  some generalized phantom cosmologies
found in \cite{chimento} in the context of k-essence cosmologies (see
expressions  (48) and (50) in that reference). If we further set $A=\la$ y
$n=-\la^4/4$ in expressions (\ref{ct}), (\ref{ptb}), (\ref{tA}), (\ref{tv0})
and (\ref{po}), then we can write those solutions in the form in which  they
were presented in \cite{Lidsey}, that is,

\be
\n{L1}
\bar A=\pm\frac{2}{\la}, \qquad \bar\phi=\pm\la\ln|t|=\pm\frac{\la^2}{2}\phi,
\ee

\no from where it is deduced that

\be
\n{L}
\bar A\bar\phi=\pm\frac{2\bar\phi}{\la}=\la\phi,
\ee

\no and finally

\be
\n{L2}
\bar V_0=\frac{\la^2}{4}(3\la^2+2),
\ee

\be
\n{L3}
\bar V=\bar V_0{\mbox e}^{-\bar A\bar\phi}=
\bar V_0{\mbox e}^{\mp 2\bar\phi/\la}=
\bar V_0{\mbox e}^{-\la\phi}.
\ee

Then, the solution presented in \cite{Lidsey} corresponds to a particular
application of the second transformation type. Actually, in his paper
\cite{Lidsey}, the author only presented models with potentials that
decrease as $\phi$ increases, which follow from taking the lower sign in
(\ref{L1}) and (\ref{L3}).

Interestingly, using the concept form-invariance, and with the help of the
parameters  $s$ y $\bar s$ we have unified the procedures already existing in
the literature, and as a byproduct we have obtained the alternative procedures
associated with the third and fourth transformations.

\section{Conclusions}

Our main goal here has been giving  a  general prescription for generating
phantom cosmologies which  unifies those obtained from the
Einstein-Klein-Gordon with both sign of the kinetic term. The algorithm
exploits form-invariance transformations, which have been discussed earlier in
relation to cosmology as part of a long-term project. After some revision of
the most important  properties of the transformations, we have concentrated on
linear transformations that multiply the Hubble factor by a negative constant
factor $n$, so that contracting solutions will be transformed into expanding
ones and {\it vice versa}. The transformation in the scale factor will be a
generalized version of the $a\to a^{-1}$ duality investigated in \cite{letter}
and appearing in the pre-big bang scenario, specifically our transformation
will consist in $a\to a^{-\vert n\vert}$. 

Let us assume $n=-1$ and that the seed solution is
the typical contracting (expanding) solution  in the $t<0$ ($t>0$) region  ending (beginning) in
a singularity at $t=0$ (see Fig. 1). Given that $a\to
a^{-1}$, the $t<0$ branch of the transformed solution will have a expanding character and it will reach a big rip at $t=0$,
thus representing a phantom cosmology. The $t>0$ branches of the transformed and seed solutions are the standard duals of each other.

For a constant $n$, the form-invariance transformations we consider can be
viewed as group of uniparametric linear transformations acting on the sets of
variables $(\rho,H,p)$ or $(\dot{\phi}^2,H,V)$, where $\dot\phi^2$ and $V$ are
the kinetic term and the potential of a conventional scalar field. Although we
fix the value of the parameter of the group $n$, some discrete symmetry
remains associated with the possibility of choosing a positive or a negative
sign in kinetic term in the effective energy density of the scalar field. In
addition, two different symmetry transformations can be constructed between
the CSF equation set and the NCSF equation set. Combining those possibilities,
we obtain four transformations that revert the sign of the barotropic index,
and two of them generate phantom cosmologies.

In conclusion, the procedures to transform conventional scalar field
cosmologies into phantom ones which existed up to now in the literature can be
viewed as an application of form-invariance transformations.

\section*{Acknowledgments}
L.P.C. is partially funded by the University of Buenos Aires  under
project X223, and the Consejo Nacional de Investigaciones Cient\'{\i}ficas y
T\'ecnicas.  R.L. is supported by  the University of the Basque Country through research grant 
UPV00172.310-14456/2002, by the Spanish Ministry of Science and Technology through research grant  BFM2001-0988, and  by the Basque Government through fellowship BFI01.412. 

\end{document}